\documentclass{aastex62}
\usepackage{graphicx}
\usepackage{dcolumn}
\usepackage{bm}

\usepackage[usenames,graphicx]{}
\usepackage{color}
\newcommand{\rt}{\color{black}\bf}
\begin{document}
\title{The growth of stellar mass black hole binaries trapped in an accretion disk of active galactic nuclei}
\author{Shu-Xu Yi}
\email{yishuxu@hku.hk}
\affiliation{Department of Physics, The University of Hong Kong}
\author{K.S. Cheng}
\email{hrspksc@hku.hk}
\affiliation{Department of Physics, The University of Hong Kong}
\author{Ronald E. Taam}
\email{taam@asiaa.sinica.edu.tw}
\affiliation{Academia Sinica Institute of Astronomy and Astrophysics, Taipei 10617, Taiwan}
\affiliation{Department of Physics \& Astronomy, Northwestern University, 2145 Sheridan Road, Evanston, IL 60208, USA}
 
\begin{abstract}
Among the four black hole binary merger events detected by LIGO, six progenitor black holes have masses greater
than 20\,$M_\odot$. The existence of such massive BHs calls for extreme metal-poor stars as the progenitors. An alternative possibility that a pair of stellar mass black holes each with mass $\sim7\,M_\odot$ increases to $>20\,M_\odot$ via accretion from a disk surrounding a super massive black hole 
in an active galactic nucleus is considered. The growth of mass of the binary and the transfer of orbital 
angular momentum to the disk accelerates the merger. Based on the recent numerical work of \cite{2017MNRAS.469.4258T}, 
it is found that, in the disk of a low mass AGN with mass $\sim10^6\,M_\odot$ and Eddington ratio $>0.01$, 
the mass of an individual BH in the binary can grow to $>20\,M_\odot$ before coalescence 
provided that accretion takes place at a rate more than 10 times the Eddington value. The mechanism predicts a new 
class of gravitational wave sources involving the merger of two extreme Kerr black holes associated with active 
galactic nuclei and a possible electromagnetic wave counterpart.
\end{abstract}
\keywords{gravitational waves, black holes, AGN}
\section{Introduction} \label{sec:intro}
Since the first detection of gravitational waves (GW) by the Laser Interferometer Gravitational-Wave 
Observatory (LIGO) in 2015, there have been four confirmed binary black hole merger events 
\citep{2016PhRvX...6d1015A,2016PhRvL.116x1103A,2017PhRvL.118v1101A,2017PhRvL.119n1101A}. Six out of the 
total eight progenitor black holes have mass $\gtrsim20\,M_\odot$, which may point to new formation mechanisms other than the traditional stellar evolutionary  channels. Specifically, \cite{2001ApJ...554..548F} theoretically 
estimated the mass distribution of BHs as the evolutionary remnant of massive stars (except for population 
III stars). The mass of BHs formed in this way can not be in excess of $\sim10-15M_\odot$, due to 
significant wind mass loss. Observations of BHs in X-ray binaries in the Galaxy are in accordance 
with a similar mass distribution \citep{2011ApJ...741..103F}. 

Low metallicity ($Z<0.1\,Z_\odot$; $Z_\odot$ is the solar metallicity) star progenitors can remain 
more massive throughout their evolution as they undergo less wind mass loss and, therefore, are likely to 
produce BHs of higher mass. It is generally believed that these low metallicity stars are formed in the 
early universe \citep{2015ApJ...806..263D,2016ApJ...819..108B}, which \textbf{seems} in contradiction with the 
low redshift of the detected GW events (three with $z\sim0.1$ and one with $z\sim0.2$). However, by taking into consideration two factors: 1. a long delay 
time from the birth of the BH binary to coalescence of $t_{\rm{merge}}\sim10\,$Gyr \citep{2014MNRAS.442.2963K}; 2. a significantly 
higher low-metallicity star formation rate at low redshifts than previously thought \citep{2016ApJ...822..108H}, this apparent conundrum is solved \citep{2016Natur.534..512B}. 

Here, we propose an alternative 
evolutionary channel in which BHs with masses greater than $20\,M_\odot$ 
can form without the requirement of a low metalicity environment. Specifically, a pair of stellar 
mass BHs are trapped by and accreted mass from an ambient accretion disk surrounding a supermassive black 
hole (SMBH) in an active galactic nucleus (AGN). There are a number of recent theoretical works studying related scenarios. In particular, models involving the formation, migration, trapping, hardening and driven merger 
of stellar mass BH binaries in an accretion disk of a SMBH have previously been studied most recently by 
\cite{2012MNRAS.425..460M,2017MNRAS.464..946S,2017ApJ...835..165B,2018MNRAS.474.5672L}. However, based on a simple description of 
the binary interaction with the disk, it was found that little mass was accreted onto the BHs during their 
inspiral and migration. Here, we consider the same process, but based on the results of the most recent 
numerical study of the binary interaction with the circumbinary disk by \cite{2017MNRAS.469.4258T}. It is 
found that there are possible circumstances under which BHs can accrete a non-negligible amount of material 
from the disk. In this case, there is no requirement for a low metallicity of the stellar population for the 
formation of high mass stellar BHs $(>20\,M_\odot)$. 

In this Letter, we describe the evolution of binary BHs in terms of a model of mass accretion and orbital 
shrinkage of a system embedded in a circumbinary accretion disk in \S 2.  With assumed distributions of initial 
parameters of the binaries, the distribution of BH mass at coalescence is determined in \S 3 and the migration 
of the binary is discussed in \S 4. Finally, we summarize and discuss the implications of our results in the 
last section. 
  
\section{Accretion onto the binary}
For simplicity, we limit our discussion to the case of an equal mass binary ($M_1=M_2=m_{\rm{BH}}\,M_\odot$. This simplification can be justified by the results of \cite{2014ApJ...783..134F}, who find that the mass ratio of a binary embedded within a disk tends to unity in a 
Keplerian circular orbit. Taking the time derivative of the expression of the orbital angular momentum $J$, and with the further assumption that each BH in the binary is accreting at the same rate, we obtain the following relation:
\begin{equation}
\frac{\dot{J}}{J}=\frac{3\dot{M}_{\rm{bin}}}{2M_{\rm{bin}}}+\frac{\dot{a}_{J}}{2a},\label{eqn:alwaystrue}
\end{equation}
where $M_{\rm{bin}}=2\,m_{\rm{BH}}\,M_\odot$ and $m_{\rm{BH}}$ is the dimensionless mass of an individual BH, $a$ 
is the orbital separation of the binary, $\dot{a}_{J}$ is the time rate of change of $a$ due to the exchange of angular 
momentum.  The numerical simulations of \cite{2017MNRAS.469.4258T}, which also take account of the dominant 
torques exerted on the binary system associated with the distortion of the mini disks surrounding the two BHs,
can be described by the following empirical relation (with physical unit restored):
\begin{equation}
\dot{J}=(-0.21\,\tau_{\rm{sink}}+0.437)\,\dot{M}_{\rm{bin}}\sqrt{GMa},\label{eqn:empirical}
\end{equation}
where $\tau_{\rm{sink}}$ is the dimensionless sink time-scale that describes the mass removal rate (due to 
accretion onto the BHs). We treat the net accretion rate onto individual BHs $\dot{m}_{\rm{BH}}$ and $\tau_{\rm{sink}}$ as independent parameters. This follows from the fact that $\tau_{\rm{sink}}$ alone does not determine the mass flow rate from the mini disk to the BH without providing the surface density of the mini disk. In addition, the net accretion rate can differ from the mass flow rate in the mini disk as result of, for example, outflows. 

Combining equations (\ref{eqn:alwaystrue}) and (\ref{eqn:empirical}), we have:
\begin{equation}
\frac{\dot{a}_J}{\dot{m}_{\rm{BH}}}=-\gamma\frac{a}{m_{\rm{BH}}},\label{eqn:amac}
\end{equation}
where $\gamma=1.68\,\tau_{\rm{sink}}-0.496$. 
\cite{2017MNRAS.469.4258T} mentioned that in any physical case $\tau_{\rm{sink}}$ should be larger than 2.1, and $\tau_{\rm{sink}}$=5 corresponds to a slow sink.

The orbital shrinkage rate due to gravitational radiation is given by 
\begin{equation}
\dot{a}_{12,\rm{GW}}=-7.88\times10^{-14}a^{-3}_{12}m^3_{\rm{BH}},
\label{eqn:gw}
\end{equation}
where $\dot{a}_{12,\rm{GW}}$ is the $\dot{a}$ due to GW radiation in unit of 10$^{12}$\,cm/yr.

For accretion onto the BH, we assume that its rate is proportional to its mass,
\begin{equation}
\dot{m}_{\rm{BH}}=\eta m_{\rm{BH}}/\tau_{\rm{acc}},
\end{equation}
where $\tau_{\rm{acc}}$ is the Eddington accretion time scale equal to $10^8$\,yr (assuming a radiative efficiency of $\sim20\%$\footnote{A different choice of the radiative efficiency is equivalent to a varies in $\eta$}), and $\eta$ is the 
Eddington ratio. 

The total change in the rate of the separation is given by $\dot{a}=\dot{a}_{\rm{GW}}+\dot{a}_J$,  where it is assumed that the individual torque contributions can be linearly added.  Although this is a common assumption, nonlinearities may result from distortions of the disk.
Combining 
equations (3, 5 and 6), and eliminating $t$, we obtain a differential equation between $a$ and $m_{\rm{BH}}$:
\begin{equation}
\frac{da_{12}}{dm_{\rm{BH}}}=-\gamma\frac{a}{m_{\rm{BH}}}-7.88\times10^{-6}a^{-3}_{12}m^2_{\rm{BH}}\eta^{-1}.
\label{eqn:totaldifferential}
\end{equation}
The solution of equation (\ref{eqn:totaldifferential}) is illustrated in Figure 1 for $\eta$ ranging from 1 to 1000 and 
$\tau_{\rm{sink}}$ ranging from 3 to 5. 
Initially, for large $a$ and small $M_{\rm{BH}}$, the shrinkage of $a$ is 
dominated by the angular momentum transfer between the binary and the circumbinary disk. Therefore, the dependence of 
$a$ on $M_{\rm{BH}}$ is given by a power law, 

\begin{equation}
a=a_0\left(\frac{m_{\rm{BH}}}{m_{\rm{BH},0}}\right)^{-\gamma},
\label{eqn:pl}
\end{equation}
where $a_0$ and $m_{\rm{BH},0}$ is the initial separation and the BH mass (dimensionless) respectively. When $a$ is 
sufficiently small, GW radiation dominates the evolution, resulting in the turn over in Fig. 1.

\begin{figure}
\centering
\includegraphics[width=8cm]{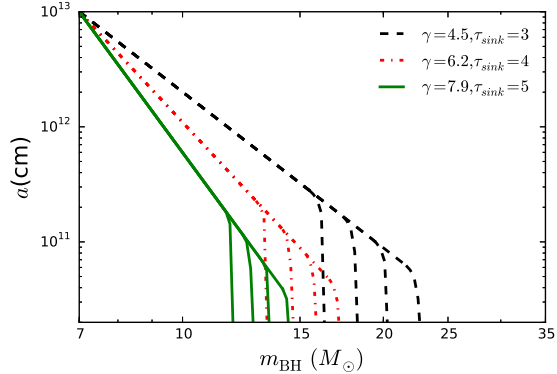}
\caption{The shrinkage of the binary orbital separation as a function of the BH mass. Curves delineated 
with different line styles and colors correspond to $\tau_{\rm{sink}}$ from 3 to 5 
($\gamma$=4.5 to 7.9). For each value of $\gamma$, four 
branches are shown corresponding to $\eta=1, 10, 100, 1000$ from left to right.}\label{fig:figure1}
\end{figure}

The final mass of the BH before coalescence can be evaluated approximately by equating the $\dot{a}_J$ and $\dot{a}_{\rm{GW}}:$
\begin{equation}
\gamma\frac{a_{12}}{m_{\rm{BH}}}=7.88\times10^{-6}a^{-3}_{12}m^2_{\rm{BH}}\eta^{-1}.
\end{equation}
Eliminating $a$ with equation (\ref{eqn:pl}), we have:
\begin{equation}
m_{\rm{BH,c}}=\left(\frac{a^4_{12,0}\eta\gamma m^{4\gamma}_0}{7.88\times10^{-6}}\right)^{1/(3+4\gamma)},\label{eqn:bhc}
\end{equation}
where $m_{\rm{BH,c}}$ and $m_0$ are the mass of the individual BH at GW dominated coalescence and the 
initial mass of the BH, both in unit of $M_\odot$; $a_{12,0}$ is the initial separation in unit of $10^{12}\,$cm. The mass evaluated with equation (9) is conservative, because the binary can continue accreting until its merger. However, the correction to $m_{\rm{BH,c}}$ will be very small, considering the short duration of the plunging phase. Figure 1 clearly shows how $a_0$, $m_{\rm{BH,0}}$, $\tau_{\rm{sink}}$ and $\eta$ determine the final mass $m_{\rm{BH,c}}$: $a_0$ is the intercept on the vertical axis when the BH has a mass of $m_{\rm{BH,0}}$; a larger value of $\tau_{\rm{sink}}$ yields a greater power law slope on the declining portion, thus resulting in a smaller $m_{\rm{BH,c}}$; a larger value of $\eta$ yields a larger $m_{\rm{BH,c}}$. It is found that the final mass $m_{\rm{BH,c}}$ is more sensitive to $\tau_{\rm{sink}}$ than to $\eta$. 

\section{Distribution of BH mass at coalescence}

To determine the population  distribution of $m_{\rm{BH,c}}$,  the distribution of each parameter on the 
right side of equation (\ref{eqn:bhc}) is required.  For definiteness, a Gaussian distribution of the initial 
mass of BHs is assumed, with the mean mass of $m_{\rm{BH},\mu}=7.8\,(M_\odot)$ and the standard deviation 
$m_{\rm{BH},\sigma}=1.2\,(M_\odot)$ \citep{2010ApJ...725.1918O,2011ApJ...741..103F}.  The parameter, $\gamma$, 
is calculated from $\tau_{\rm{sink}}$, the distribution of which is assumed to be uniform between $\tau_{\rm{sink}}=3$ to $\tau_{\rm{sink}}=5$ 
Finally, the Eddington ratio, $\eta$, is assumed to range 
from $1$ to $1000$ log-uniformly. 

The actual distribution of the orbital separation, $a$, of the binary BH system is uncertain.  However the 
Roche radius of the binary BH in orbit about the SMBH should provide an upper limit for $a$, for otherwise 
any binary would be torn apart by the tidal force of the SMBH. The Roche radius $R_{\rm{R}}$ is approximated by:
\begin{equation}
a_{\rm{0,max}}=R_{\rm{R}}=\left(\frac{2m_{\rm{BH}}}{m_{\rm{SMBH}}}\right)^{1/3}R_0,
\end{equation}
where $R_0$ is the initial distance from the binary to the SMBH. If we take the mass of the 
supermassive BH (in unit of $M_\odot$) $m_{\rm{SMBH}}\sim 10^6$, 
$2\,m_{\rm{BH}}\sim 10$, $R_0\sim0.01\,$pc, then $R_{\rm{R}}\sim 6 \times 10^{14}\,$cm. 
Stellar interactions with 
the binary BH can affect the orbital separation distribution and \cite{2017ApJ...835..165B} considered the ionization 
of the binary by softening encounters with stars. In this context, the time scale of the ionization can be estimated by:
\begin{equation}
t_{\rm{ion}}\approx10\times\frac{M_{\rm{SMBH}}}{M_{\rm{bin}}}t_{\rm{orb}},
\end{equation}
where $M_{\rm{SMBH}}$ is the mass of the SMBH and $t_{\rm{orb}}$ is the orbital period of the binary around the 
SMBH. For $M_{\rm{SMBH}}>10^6\,M_\odot$, $M_{\rm{bin}}=20\,M_\odot$ and $R=0.01$\,pc, $t_{\rm{ion}}>5\times10^7$\, 
yr, which is orders of magnitude greater than the merger time scale. Therefore, 
we can safely ignore this process.  

A flat distribution of $a$ in log space is used, i.e., \"Opik's law 
\citep{2002A&A...382...92S,2007AJ....133..889L,2007ApJ...670..747K} with a large $a$ cut-off 
as the distribution of $a$. There should also be a cut-off at the small $a$, where the GW radiation dominates at 
the outset. This value can be found by setting $m_{\rm{BH,c}}=m_0$ in equation (\ref{eqn:bhc}):
\begin{equation}
a_{12,0,\rm{min}}=\left(7.88\times10^{-6}\eta^{-1}\gamma^{-1}m^3_0\right)^{1/4}.
\end{equation}

The resulting population distribution of $M_{\rm{BH,c}}$ for a range of $a$ cut-off values is 
plotted in Fig. 2. The distribution 
of BH mass peaks at $\sim10\,M_\odot$, which is independent of the large $a$ cut-off, although a larger 
cut-off extends the tail of the distribution to higher masses. In both distributions, there is a 
significant fraction of BHs that can increase its mass to greater than $20\,M_\odot$. 

\begin{figure}
\centering
\includegraphics[width=8cm]{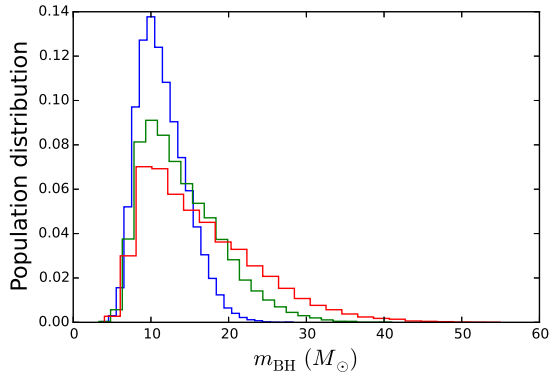}
\caption{The population distribution of $m_{\rm{BH,c}}$. The blue histogram corresponds to the large 
$a$ cut-off at $10^{13}$\,cm, the green histogram corresponds to the large $a$ cut-off at $10^{14}$\,cm and 
the red histogram corresponds to the cut-off at $6\times10^{14}$\,cm.}
\end{figure}

\section{Migration to SMBH}
Since the binary is trapped in the accretion disk of an AGN, it will migrate closer to the SMBH due to 
its interaction.  If the orbital separation of the double black hole binary system is much less than 
the scale height of the disk, the binary can be approximated by a massive point object. We use the disk-satellite 
interaction to estimate the time scale of migration. Specifically, \cite{2011ApJ...726...28B} have shown 
that if the object is sufficiently massive to open a gap in the disk the horseshoe drag is much reduced and 
the Lindblad torque balances the viscous torque exerted by the disk (type II migration). The gap-opening criterion is \citep{2011ApJ...726...28B}:
\begin{equation}
\frac{3H}{4R_0}(q/3)^{-1/3}+50\alpha\left(\frac{H}{R_0}\right)^2/q<1. 
\end{equation}
For a SMBH with 
$\sim10^6\,M_\odot$, a binary BH with mass larger than $20\,M_\odot$ (with $\alpha=0.01$) meets the criterion of type II migration. 
In this case, the migration time scale is estimated with the equation (6) of \cite{2011ApJ...726...28B}. We adopt $m_{\rm{SMBH}}=10^6$, and use the standard 
gas pressure dominated disk model \citep{1973A&A....24..337S} $H/R_0\approx3\times10^{-3}$ with Kramer's opacity, 
$\alpha=0.01$ and $r_0=0.01$\,pc. This yields $\tau_{\rm{M}} \approx1.1\times10^8$\,yrs. Hence, for an 
Eddington ratio $\eta>\alpha/10^{-2}$, the migration time scale does not limit the growth of the BH mass via 
accretion for low mass AGNs. For AGN with $m_{\rm{SMBH}}>10^8$, the migration will be type I and the migration 
time is $\sim10^6\,\text{yr}$ \citep{2011ApJ...726...28B}, which is be too short for a significant mass increase of the binary.

\section{Conclusion and discussion}

The nuclei of galaxies are known to be regions of high stellar density.  The star formation rates can be high 
and the migration of massive objects towards the central region effective.  Hence, it is possible that a large 
population of stellar mass BHs could be harbored in this region, and some of them might be trapped in the gaseous 
disk around the AGN \citep{2017ApJ...835..165B}. We show in this Letter that, given a super-Eddington accretion 
rate ($\eta$ in range between $\sim10$ and 1000), stellar mass binary BHs can accrete significant amount 
of mass from a circumbinary disk before the GW radiation dominates the coalescence phase with a non-negligible 
fraction growing to $>20\,M_\odot$. The above mentioned evolutionary channel serves as an alternative pathway to 
produce the unusually massive stellar mass BH binaries detected by LIGO.   
\subsection{The validity of the circular orbit}
The binary orbit in a hierarchical triplet may evolve to a large eccentricity due to the Kozai-Lidov mechanism \cite{1962AJ.....67..591K,1962P&SS....9..719L}. The first nonzero term of the Kozai-Lidov mechanism is the quadrupole term, which vanishes when the inclination of the inner orbit is smaller than a critical value of $\sim$39.2$^\circ$. In the next order, i.e., the octupole Kozai-Lidov mechanism, a nearly coplanar inner orbit with very small initial eccentricity will not evolve to a large eccentricity \citep{2003ApJ...592.1201L,2014ApJ...785..116L}. Therefore we limit our discussion to an initially coplanar and 
circular orbit, which is stable and self-consistent.
\subsection{Accretion rate onto the binary}
Here, we examine the consistency for a super-Eddington accretion rate corresponding to $\eta=1-1000$, given the 
densities in the AGN disk. An upper limit of the accretion rate onto the binary can be estimated with (as in equation (2) of \cite{2017MNRAS.464..946S}):
\begin{equation}
\dot{M}=\pi\rho\sigma_{\rm{gas}}r_{\rm{acc}}\text{min}[r_{\rm{acc}},H],
\end{equation}
where $\rho$ and $H$ are the density and scale height of the AGN disk respectively, $r_{\rm{acc}}$ is the radius of accretion:
\begin{equation}
r_{\rm{acc}}=\frac{GM_{\rm{bin}}}{\sigma^2_{\rm{gas}}},
\end{equation}
where $\sigma^2$ is the squared sum of three terms: 1. the sound speed of the gas $c_{\rm{s}}$, 2. the velocity shear across the Hill radius of the binary and 3. the relative velocity between the binary and the disk due to eccentric and inclined orbit (see \cite{2017MNRAS.464..946S}). In this paper, we consider the binaries in a coplanar and circular orbit around the SMB and, therefore, the third term in $\sigma^2_{\rm{gas}}$ vanishes. $\rho$ can be estimated with the formula of a standard thin disk:  
\begin{widetext}
\begin{equation}
\rho=5\times10^{-11}\left(\frac{\alpha}{0.01}\right)^{-7/10}\eta^{11/20}_{\rm{SMBH}}\left(\frac{M_{\rm{SMBH}}}{10^6\,M_\odot}\right)^{47/40}\left(\frac{R_0}{0.01\,\rm{pc}}\right)^{-15/8}\,\rm{g/cm^3},
\end{equation}
\end{widetext}
where $\eta_{\rm{SMBH}}$ is the Eddington ratio of the SMBH. For 
$M_{\rm{SMBH}}\sim10^6\,M_\odot$, $R_0=0.01\,\rm{pc}$, $\alpha=0.01$ and $M_{\rm{bin}}=20\,M_\odot$,  
we find that $\dot{M}$ is larger than $10^3$ times of the Eddington rate of the binary as long as $\eta_{\rm{SMBH}}>
0.01$. 
When the orbit of the binary is retrograde with respect to the accretion disk of the AGN, the relative velocity between the 
binary and the gas in the accretion disk dominants $\sigma^2_{\rm{gas}}$. As a result, the 
mass inflow rate is much smaller and insufficient to support the required accretion rate. Therefore, we limit our 
discussion in this paper to the case of a prograde orbit. 
In the case of super-Eddington accretion, the actual accretion rate onto the BH could be only a fraction of the 
total matter inflow rate, because substantial mass loss in an outflow is expected (see recent works of 
\cite{2014ApJ...780...79Y} and \cite{2014ApJ...796..106J} for instance). Hence, the accretion rate in this paper 
refers to the net value onto the individual BH. $R_0=0.01\,\rm{pc}$ is taken as a reference value in the above discussion. For larger radii ($R_0>0.1\,\rm{pc}$), the AGN disk is no longer stable under its self-gravity and, thus, star formation becomes important \citep{2017MNRAS.464..946S}. As a consequence, the physical picture {\rt might need modifications.}

\subsection{Observational consequences}

In our channel, stellar mass BHs can accrete several times their initial mass from the accretion disk. We 
thus expect that the resulting BHs can become extreme Kerr BHs before coalescence with their spin axes 
aligned. For such 
systems, the results of numerical simulations \citep{2014PhRvD..90j4004H} indicate that the total energy loss in
GW radiation would be $\sim10\%$ of the rest mass energy or 2-3 times higher than in the case for a random 
orientation.  Although the currently detected GW events do not provide support for such progenitors, future 
events may be found which correspond to rapidly spinning and aligned BHs, a likely consequence of our model. 

A natural expectation in this formation channel of such GW sources is that they are associated with AGNs. The super-Eddington accretion onto the BHB can give rise to luminous X-ray emission, which may outshine a sub-Eddington accreting AGN in the same wavelength band (see \citealt{2017ApJ...835..165B} and \citealt{2017MNRAS.464..946S}). Furthermore, after the coalescence of the BHB, remnant matter from the mini disk may fall back onto the newly formed Kerr BH at an even higher rate. Such a possibility may trigger the formation of a relativistic jet through the Blandford-Znajek mechanism \citep{1977MNRAS.179..433B}. As pointed out by \cite{2017ApJ...835..165B}, gamma ray burst-like counterpart could be produced in this case. Besides the radiation at high energies, the jet might also give rise to coherent radio emission, which will be the topic of a future paper.
\acknowledgments
The authors appreciate the suggestions from reviewers which improved the manuscript a lot. KSC and SXY are supported by a GRF grant under 17310916.

\end{document}